\documentclass[prl, superscriptaddress, reprint,showpacs,amsmath,amssymb,floatfix]{revtex4-1}

\usepackage[inline]{asymptote}
\usepackage[alsoload=synchem]{siunitx}
\DeclareSIUnit{\cal}{cal} 
\DeclareSIUnit{\Newton}{N}

\newcommand{\mm}{\textcolor{black}}
\newcommand{\eg}{{\textit {e.g.~}}} 
\newcommand{\ie}{{\textit {i.e.~}}} 
 \usepackage{hyperref}
 \hypersetup{colorlinks, citecolor={blue},linkcolor={cyan}, urlcolor={blue},
 pdftitle={SAGE},
 pdfauthor={M Mosayebi}, pdfdisplaydoctitle}

\begin{document}

\title{Beyond icosahedral symmetry in packings of proteins in spherical shells}

\author{Majid Mosayebi}
\email{majid.mosayebi@bristol.ac.uk}
\affiliation{School of Mathematics, University of Bristol, University Walk, Bristol BS8 1TW, UK}
\affiliation{BrisSynBio, Life Sciences Building, Tyndall Avenue, Bristol BS8 1TQ, UK}
\author{Deborah K. Shoemark}
\affiliation{BrisSynBio, Life Sciences Building, Tyndall Avenue, Bristol BS8 1TQ, UK}
\affiliation{School of Biochemistry, University of Bristol, Bristol BS8 1TD, UK}
\author{Jordan M. Fletcher}
\affiliation{School of Chemistry, University of Bristol, Cantock’s Close, Bristol BS8 1TS, UK}
\author{Richard B. Sessions}
\affiliation{BrisSynBio, Life Sciences Building, Tyndall Avenue, Bristol BS8 1TQ, UK}
\affiliation{School of Biochemistry, University of Bristol, Bristol BS8 1TD, UK}
\author{Noah Linden}
\email{n.linden@bristol.ac.uk}
\affiliation{School of Mathematics, University of Bristol, University Walk, Bristol BS8 1TW, UK}
\author{Derek N. Woolfson}
\email{d.n.woolfson@bristol.ac.uk}
\affiliation{BrisSynBio, Life Sciences Building, Tyndall Avenue, Bristol BS8 1TQ, UK}
\affiliation{School of Biochemistry, University of Bristol, Bristol BS8 1TD, UK}
\affiliation{School of Chemistry, University of Bristol, Cantock’s Close, Bristol BS8 1TS, UK}
\author{Tanniemola B. Liverpool}
\email{t.liverpool@bristol.ac.uk}
\affiliation{School of Mathematics, University of Bristol, University Walk, Bristol BS8 1TW, UK}
\affiliation{BrisSynBio, Life Sciences Building, Tyndall Avenue, Bristol BS8 1TQ, UK}

\onecolumngrid

\begin{abstract}
The formation of quasi-spherical cages from protein building blocks is a remarkable self-assembly process in many natural systems, where a small number of elementary building blocks are assembled to build a highly symmetric icosahedral cage. In turn, this has inspired synthetic biologists to design \textit{de novo} protein cages. We use simple models, on multiple scales, to investigate the self-assembly of a spherical cage, focusing on the regularity of the packing of protein-like objects on the surface. Using building blocks, which are able to pack with icosahedral symmetry, we examine how stable these highly symmetric structures are to perturbations that may arise from the interplay between flexibility of the interacting blocks and entropic effects. We find that, in the presence of those perturbations, icosahedral packing is not the most stable arrangement for a wide range of parameters; rather disordered structures are found to be the most stable. Our results suggest that (i) many designed, or even natural, protein cages may not be regular in the presence of those perturbations, and (ii) that optimizing those flexibilities can be a possible design strategy to obtain regular synthetic cages with full control over their surface properties.
\end{abstract}

\maketitle

\twocolumngrid


Many examples of self-assembled quasi-spherical shells or cages are found in biology. Small ferritin cages, numerous viral capsids, clathrin, and large carboxysomes in bacteria are all cages composed of protein sub-units, and the resulting structures are used for packaging and transport \cite{ferritin, caspar1962physical,Edeling:2006ga, carboxysome}. With a few exceptions, these spherical cages typically have highly ordered structures with icosahedrally symmetric shells \cite{caspar1962physical,Zandi_pnas, Bruinsma_prl, capsid_mech, kirchhausen2014,carboxysome, Wagner_Zandi}, characterized by 6 five-fold, 10 three-fold and 15 two-fold symmetry axes. Biology, through evolution, has developed very efficient routes to make icosahedral cages using a small number of elementary protein building block types. These elementary building blocks seem to assemble in a hierarchical manner \cite{Reguera,Hagan_review, hierarchical_assembly,Corchero_BMC}. They firstly assemble to make larger assemblies (hexagons and pentagons) and subsequently a spherical cage is formed from these oligomers. 

Inspired by biology, there have been several attempts to design synthetic protein cages either by taking protein engineering or \textit{de novo} design approaches \cite{Fletcher:2013fx, King:2014cf, Indelicato:2016bf, Hsia:2016ga} or by re-engineering natural protein cages \cite{Sasaki_jpcb}. In some synthetic-cage assemblies, the aim is to form a mono-dispersed, highly ordered cage \cite{Hsia:2016ga}, while in others, the goal is to employ simpler and more chemically accessible design rules to form spherical cages that are not necessarily symmetric \cite{Fletcher:2013fx}. Synthetic cages could potentially be used for many applications such as targeted drug delivery, vaccine design, nano-reactors and synthetic biology \cite{kostiainen,Dragnea_magnetic_capsid,Kushnir,Cornelissen,uchida2007,putri2015}. 

A better understanding of the assembly mechanisms in protein cages would make it possible efficiently to control the structural properties of the protein cage to best suit the particular application required. In that direction, a number of questions arise: What are the key design rules for synthetic self-assembly aiming for regular cages? How robust is the icosahedral symmetry of the protein cage? Can stable cages be constructed without this symmetry? 

By combining several coarse-grained models on multiple scales, here we investigate the robustness of the icosahedral cage against structural imperfections, arising from the flexibility of the protein building blocks, which could possibly occur in synthetic self-assembly pathways. To fix our ideas with a concrete example of a synthetic cage, we seek inspiration from the recently designed self-assembled cages (SAGEs) \cite{Fletcher:2013fx}, which use \textit{de novo} designed coiled coil (CC) peptides as building blocks.

\section*{Self-assembled cage from CC peptides}
\begin{figure}[t]
  \centering
  \includegraphics[width=\linewidth]{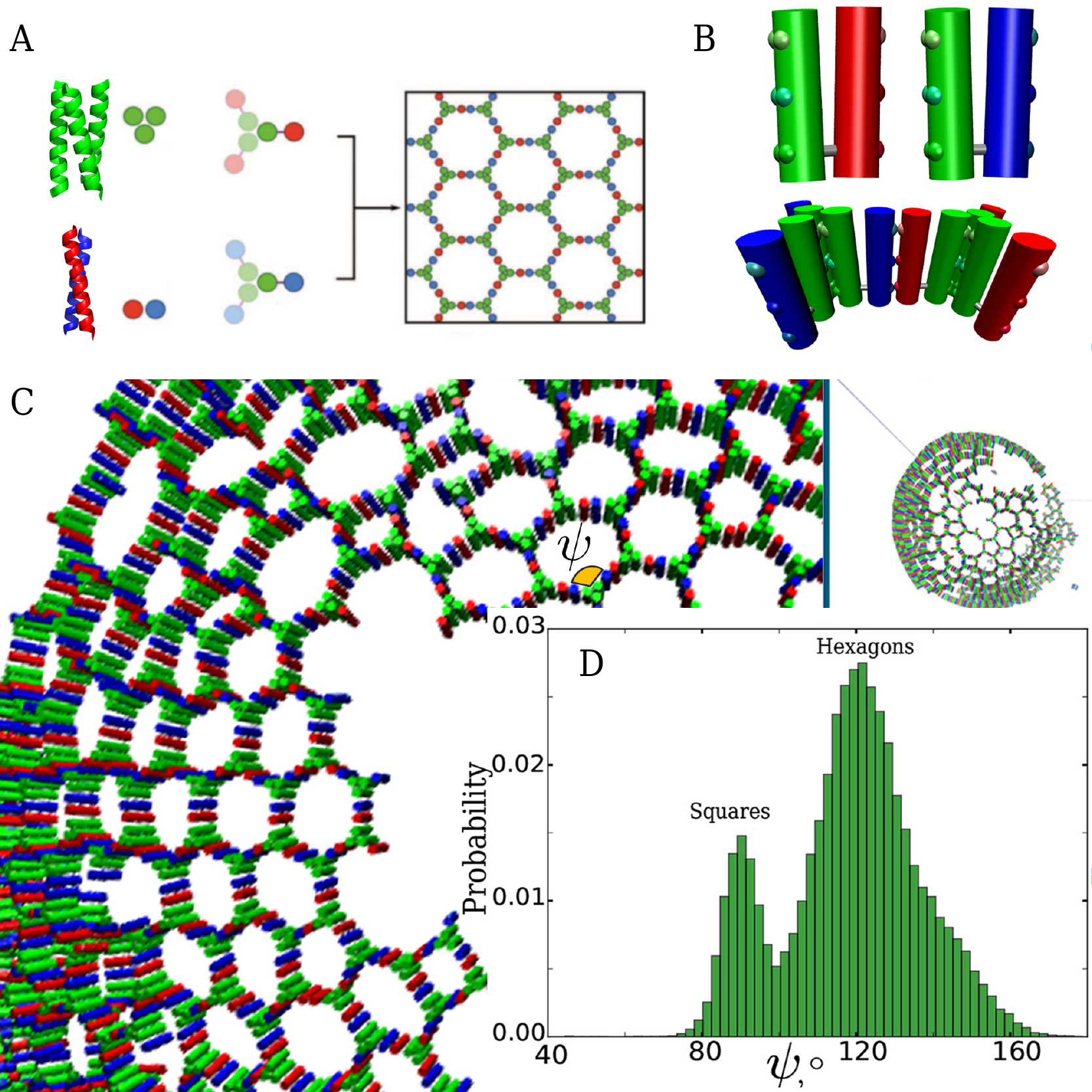}
  \caption{Self-assembled peptide cages (SAGEs) from \textit{de novo} coiled-coil (CC) peptides. \textbf{A}. Homotrimeric CC (green) and heterodimeric CC (acidic CC in red and basic CC in blue) bundles. Each CC is made from 21 amino acids with a height of 3\,nm. Two hub types are made by linking 3 homotrimeric acidic or basic CCs to a trimer bundle using 3 disulfide bonds. These hubs, when mixed, form a honeycomb lattice that closes due to the intrinsic splay between hub pairs \cite{Fletcher:2013fx}. \textbf{B}. Two types of molecules in the CC-level CG model of SAGEs (top). Attractive LJ patches are illustrated as small spheres that drive formation of trimers and dimers (bottom). \textbf{C}. Snapshots of a partially assembled SAGE obtained after annealing a mixture of pre-formed hubs. \textbf{D}. The probability distribution of the polygon angles $\psi$ averaged over the simulation trajectory.}
  \label{fig:motivation}
\end{figure}

The SAGE design comprises two, noncovalent, heterodimeric and homotrimeric CC bundles. These are joined back to back using disulfide bonds to make two complementary hubs, which when mixed form honeycomb networks (see Fig.\,\ref{fig:motivation}-\textbf{A}). The honeycomb network folds due to the intrinsic splay inherent within the designed hubs, and leads to cages with a typical diameter of \SI{100}{\nano \meter}. Thus, SAGE assembly is reminiscent of the formation of clathrin, where hexagonal and pentagonal sub-units are allowed to form due to the finite flexibility of the triskelia \cite{kirchhausen2014,Kirchhausen,Avinoam_clathrin}. The melting temperature of the trimeric CC bundle (green CCs in Fig.\,\ref{fig:motivation}-\textbf{A}) is higher than the melting temperature of the heterodimeric bundles, resulting effectively in a one-structural-unit assembly with a hub, which comes in two types depending on the type of heterodimeric CCs linked to the trimer. Hexagonal sub-units formed from 6 flat hubs would tile a flat surface.  One way of covering a spherical surface in a regular manner, without any holes, is to include 12 pentagonal sub-units, hence giving the required Euler-characteristic of 2. This strategy could be achieved by making use of the flexibility of the angle and bond potentials between hub pairs to allow the formation of the required number of pentagonal units within the honeycomb network. However, allowing more flexible or less angular specific interactions would enhance formation of other polygonal units in addition to the required number of pentagons and hexagons. Such polygons act as additional structural defects, and can potentially remove the icosahedral symmetry of the cage. This competition is a natural place for one to look for ways to optimize the cage structure.  

Studying the self-assembly process of a system as large as a SAGE over the timescale that it forms in the test-tube (up to minutes) is not feasible with atomistic detail. Thus, Coarse-Grained (CG) modeling \cite{Zandi_pnas,useful_scars,vacha,Briels,Wilber,Hagan_review,peptoid, glotzer2007, glotzer_cone, Chakrabarti, Vernizzi_pnas, Vernizzi_pnas2} is required to answer relevant questions about the dynamics of the self-assembly as well as the stability and uniqueness of the final structure. We employed a CC-level CG model of SAGEs to simulate directly the self-assembly of hubs. The overall angular specificity of the hub pairs in the CC-level CG model arises from the finite range of attractive interactions between the patches and also from the stiffness of the permanent bond. To facilitate formation of honeycomb lattice in numerical simulations, the angular specificity of the hub pairs was tuned to be more than the observed angular specificity in the atomistic simulations (See SI).

\subsection*{Assembly of the peptide network incorporates defects}
Our simulation results indicate that, even with our more angular specific model, the occurrence of non-hexagonal defects on the self-assembly pathway is likely, as it can be observed from the abundance of non-hexagonal subunits in Fig.\,\ref{fig:motivation}-\textbf{C} (For instance, see the pronounced peak for the squares in Fig.\,\ref{fig:motivation}-\textbf{D}). In addition, the lifetime of those defect-rich structures is longer than the accessible timescale of our simulations ($\approx$ \SI{100}{\mu \second}), suggesting that their contribution might not be negligible and they might significantly alter the free-energy landscape of the system. However, even at this level of coarse-graining, calculating the free-energy of a complete SAGE was not feasible due to computational limitations. We note that, consistent with our CG simulation results, the existence of non-hexagonal sub-units on the SAGE surface has recently been confirmed experimentally in the AFM measurements of the silica-coated SAGEs \cite{Jo_silica_SAGE}.

\begin{figure}
  \centering
  \includegraphics[width=\linewidth]{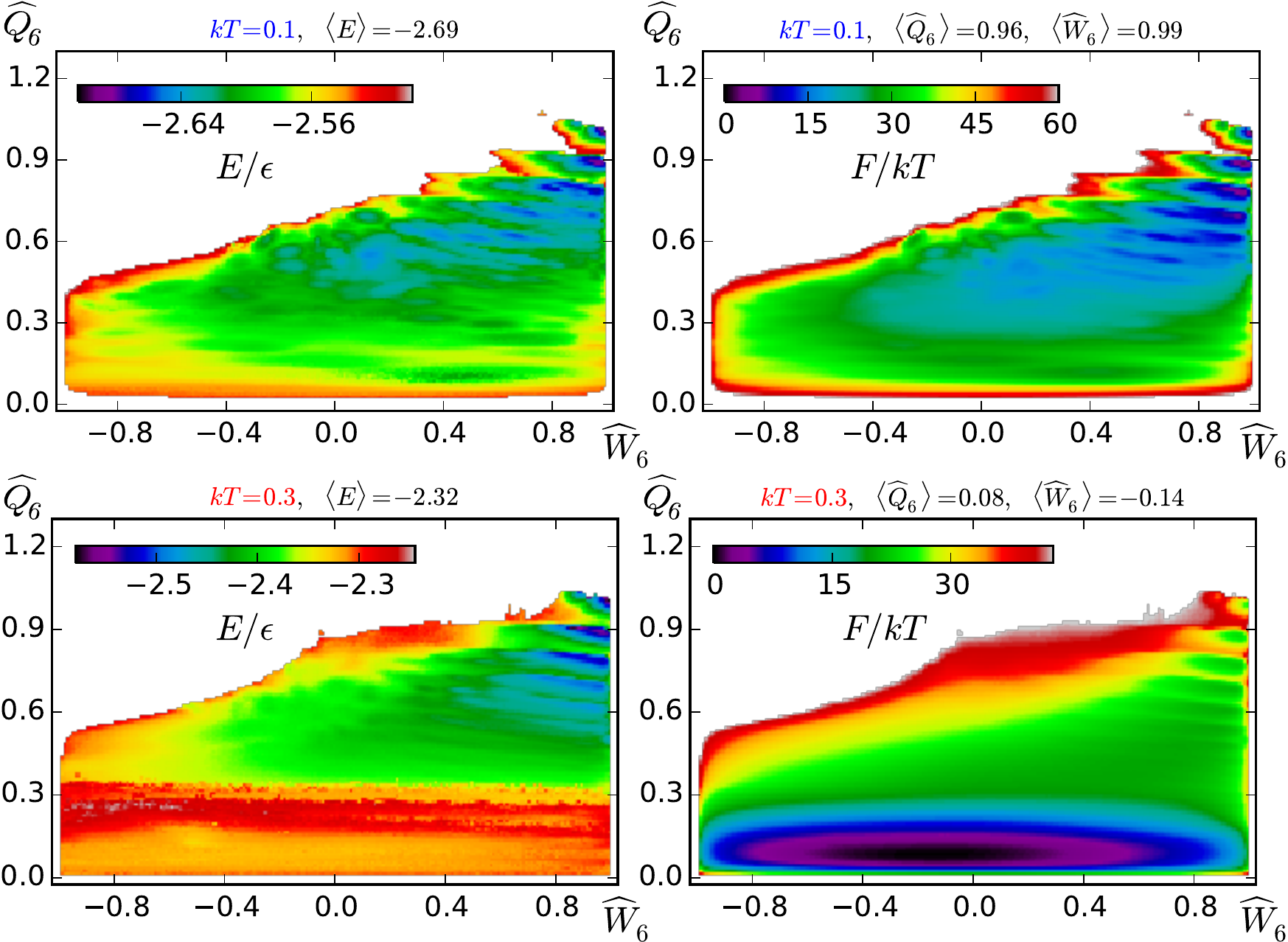}
  \caption{Polygon-level CG model energy, $E$ (left) and free-energy, $F$ (right) landscapes at $kT=0.1 \epsilon$ 
(top) and  at $kT=0.3 \epsilon$ (bottom) for an ideal spherical packing (radius $R=1.603\sigma$ corresponding to an icosahedral packing with $N=32$) composed of only hexagonal and pentagonal particles. The ratio of activities is $z_P/z_H=\exp(\Delta \mu /kT) = 12/20$. The colors vary from purple (small values) to red (large values).
}
  \label{fig:2DF}
\end{figure}

\begin{figure*}[t]
\newcolumntype{R}[1]{>{\let\newline\\\arraybackslash\hspace{0pt}}m{#1}}
\newcommand*{\Resize}[2]{\resizebox{#1}{!}{$#2$}}
\begin{center}
\begin{tabular}{R{3.26cm}   R{3.26cm}   R{3.26cm}   R{3.26cm}  R{3.26cm} }
\multicolumn{5}{c}{
\includegraphics[width=0.85\linewidth]{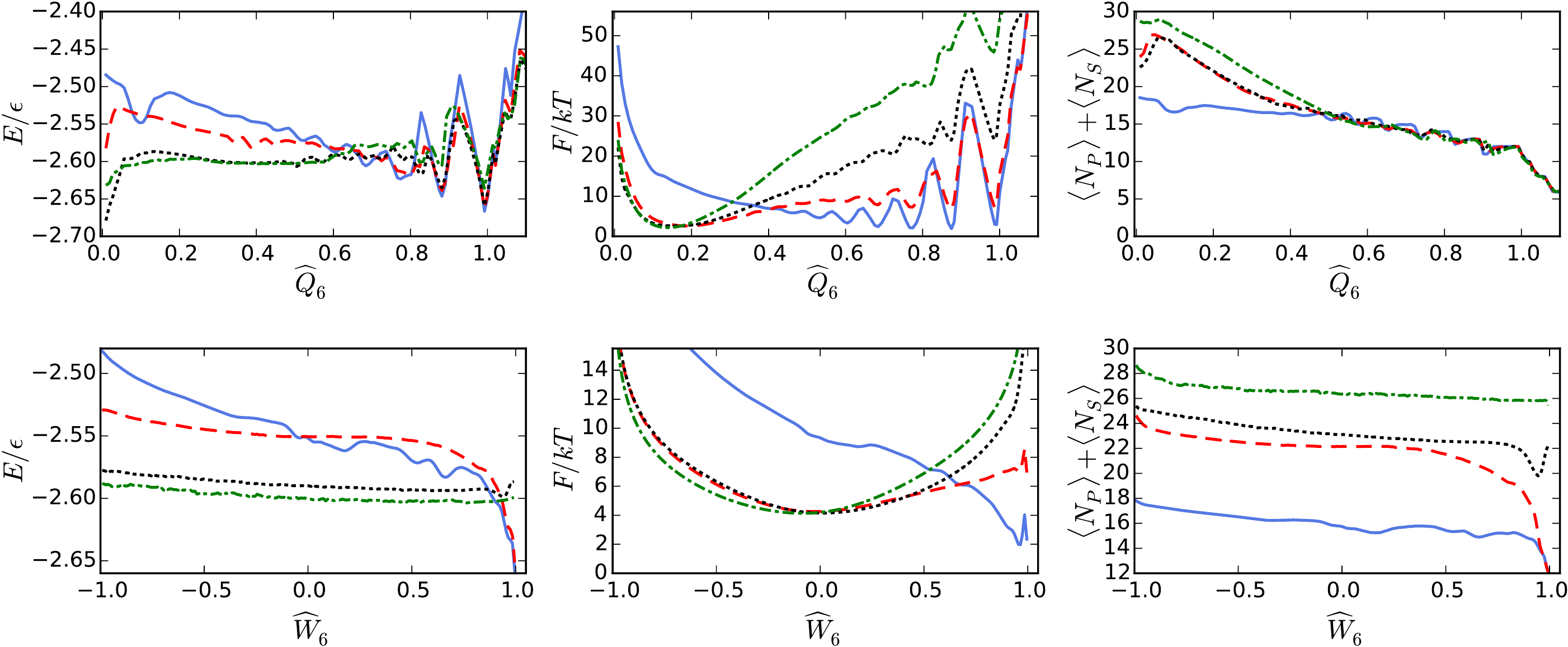} }\\
\includegraphics[width=0.85\linewidth]{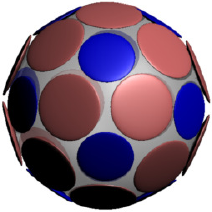} 
&
\includegraphics[width=0.85\linewidth]{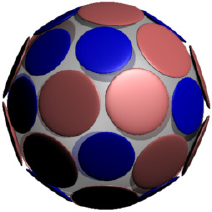} 
&
\includegraphics[width=0.85\linewidth]{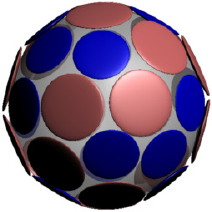} 
&
\includegraphics[width=0.85\linewidth]{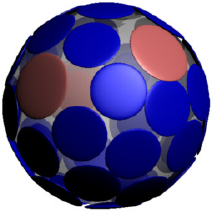} 
&
\includegraphics[width=0.85\linewidth]{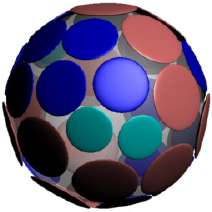} 
\\

\Resize{2.5cm}{\mathbf{i})~~  \widehat{Q}_6 \approx 1.0,~\widehat{W}_6 \approx 1.0}  &
\Resize{2.7cm}{\mathbf{ii})~~ \widehat{Q}_6= 0.87,~\widehat{W}_6 = 0.98}  &
\Resize{2.7cm}{\mathbf{iii})~~\widehat{Q}_6= 0.77,~\widehat{W}_6 = 0.95}  &
\Resize{2.7cm}{\mathbf{iv})~~ \widehat{Q}_6= 0.08,~\widehat{W}_6 = -0.1}  &
\Resize{2.7cm}{\mathbf{v})~~  \widehat{Q}_6= 0.17,~\widehat{W}_6 = 0.07}  \\ 
\Resize{2.6cm}{~~~~~N_H=20, N_P=12} &
\Resize{2.6cm}{~~~~~N_H=19, N_P=13} &
\Resize{2.6cm}{~~~~~N_H=18, N_P=14} &
\Resize{2.6cm}{~~~~~N_H=3, N_P=35} &
\Resize{3.1cm}{~~~N_H=14, N_P=17, N_S=3} 
\end{tabular}
  \caption{Energy $E$, free-energy $F$ and the number of non-hexagonal particles as a function of normalized BOOs $\widehat{W}_6$ and $\widehat{Q}_6$ at $kT=0.15 \epsilon$ and $R=1.603\sigma$ for an ideal system composed of hexagons and pentagons (solid lines) and perturbed systems composed of hexagons, pentagons and squares with varying strength for the perturbation controlled by the relative activity of the squares; weak ($z_S/z_P=0.01$, dashed lines), intermediate ($z_S/z_P=0.1$, dotted lines) and strong ($z_S/z_P=1$, dashed-dotted lines). 5 different configurations are also shown in the bottom panel:  \textbf{i}--\textbf{iv} are typical ideal packings at $kT=0.15 \epsilon$ (\textbf{i}--\textbf{iii}) and $kT=0.3 \epsilon$ (\textbf{iv}), and \textbf{v} is a typical perturbed packing at $kT=0.15 \epsilon$ and $z_S/z_P=0.01$. Copper, blue and cyan particles represent hexagons, pentagons and squares respectively. 
  }
  \label{fig:perturb}
\end{center}  
\end{figure*}

\section*{Packing of proteins on spherical shells; a mesoscale model}
To gain further insight on whether the system is able to escape from such long-lived metastable defected states and find its most symmetrical configuration, and also to measure the thermodynamic stability of the icosahedral cage we used a mesoscale model. This was on the level of polygonal units on the SAGE surface. Following the work of Zandi \textit{et al.} \cite{Zandi_pnas}, who introduce a minimal model for virus capsids, we extended the model by considering different type polygons as \mm{Lennard-Jones (LJ) particles} with different diameters $\{\sigma_i\}$, $i\in\{H, P, S, ...\}$ ($H$, $P$ and $S$ stand for Hexagon, Pentagon and Square, respectively.) that are allowed to move on the surface of the sphere or change type, while interacting via a truncated LJ interaction of strength $\epsilon$. The packing of particles on spherical surfaces have been used also as a model for colloidosomes \cite{Dinsmore_Colloidosomes, Meng_Colloidosomes, Fantoni, Burke, Bowick_prb, paquay2017}. For reference, we take our \emph{ideal}\footnote{We use the word \emph{ideal} to refer to a system that is only composed of hexagons and pentagons.} system as only composed of hexagonal ($\sigma_{H} = \sigma$) and pentagonal ($\sigma_{P} = \sigma /(2\sin(\pi/5))$) particles. The value of $\sigma_P$ was chosen such that pentagons have the same length per edge as hexagons. \mm{Therefore, a polygonal particle with $n$ edges would prefer (energetically), on average, to have $n$ neighbors. However, at the finite temperatures of our simulations, the system is allowed (by paying an energy penalty) to adopt a configuration in which the number of neighbors deviates from $n$.} It has been shown that this minimal model has the lowest free energies for packings with special numbers of particles \eg $N=12,32,42, ...$ (or equivalently special radii) corresponding to icosahedral arrangements with $N-12$ hexagons and $12$ pentagons \cite{Zandi_pnas,Paquay}, and consistent with Casper-Klug \emph{quasi-equivalence} principle for spherical virus capsids \cite{caspar1962physical}. We extended this model to include structural defects that may exist in the form of other polygonal units on the cage surface. In particular, we considered a \emph{perturbed} system that also had squares  ($\sigma_{S} = \sigma / (2\sin(\pi/4)) $). We carried out Monte Carlo (MC) simulations in the Grand-Canonical ensemble at a fixed sphere radius $R$, temperature $T$ and chemical potentials $\{\mu_i\}$.

We characterized the icosahedral symmetry of the particle packings using Steinhardt's bond orientational order (BOO) parameters $Q_6$ and $W_6$, originally defined to distinguish the local crystallinity classes in liquids and glasses \cite{steinhardt_BOO,lechner} (see Materials and Methods). BOO are rotational invariants based on spherical harmonics and are calculated from orientation of non-hexagonal particles with respect to the center of the spherical packing with values of $\{Q_6^{\rm icos}, W_6^{\rm icos}\} = \{0.6633,-0.16975\}$ for an icosahedral packing (packing \textbf{i} in Fig.\,\ref{fig:perturb}) and $\{Q_6^{\rm rand}, W_6^{\rm rand}\} \approx \{0,0\}$ for a totally random packing. We measured the free-energy profiles as a function of the normalized BOOs $\widehat{Q}_6 = Q_6/Q_6^{\rm icos}$ and $\widehat{W}_6 = W_6/W_6^{\rm icos}$, using umbrella sampling \cite{US} to facilitate the measurement of less favorable states. The umbrella weights were adjusted iteratively to have a uniform sampling as a function of BOOs. The starting configuration at each stage was chosen to be the final configuration from the previous iteration. At each temperature, we ran from 20 to 40 independent simulations, each with approximately $3 \times 10^9$ MC steps per particle and obtained the resulting profiles using the weighted histogram analysis method \cite{wham}. This procedure is repeated until the obtained free-energy profiles are converged. We note that, at the lowest temperatures studied here, the identity-exchange MC moves were mostly rejected and the dynamics became very slow.

\subsection*{Mesoscale-model predictions}
We observed that the lowest energy configurations lay at regions where the normalized BOOs were close to 1 (left panels in Fig.\,\ref{fig:2DF}). These are icosahedrally ordered arrangements that the system typically samples at low temperatures where entropic effects are negligible. Non-icosahedral arrangements with even lower energies than icosahedral arrangement have been described by authors in Ref.\,\cite{Paquay} for spherical packings composed of same-sized LJ particles. In such low-energy arrangements, particles that do not have 6 neighbors tend to cluster together. For the two system sizes that we studied here ($R=1.603\sigma$ and $R=2.469\sigma$), the optimum packings with clustering, which are typically invariant against a smaller number of symmetry operations as compared to the icosahedral packing (Fig.\,S6
-\textbf{C}), were local energy minima (with $E > E_{\rm icos}$). This indicates that icosahedral symmetry is a more robust feature of our ideal binary-mixture packings. However, for much larger packings with 312 particles, we saw arrangements with $E < E_{\rm icos}$ in which pentagons cluster together in $12\times6$ groups (see SI, Fig.\,S6
). For the ideal system, we found the icosahedral packing to be the most probable arrangement at low temperatures ($kT\lesssim0.2 \epsilon$), in agreement with previous studies \cite{Zandi_pnas,Paquay} (top-right panel of Fig.\,\ref{fig:2DF}). At sufficiently large temperatures ($kT\gtrsim0.2 \epsilon$), the entropy term becomes the dominant term in the free energy and thus, at the expense of energy, the system explores conformations with a larger number of pentagons (defects). Therefore, in the high-temperature regime, the free-energy minimum of the system is in the disordered region with small $\widehat{Q}_6$ and $|\widehat{W}_6|$ (bottom-right panel of Fig.\,\ref{fig:2DF}).

\begin{figure}
  \centering
  \includegraphics[width=\linewidth]{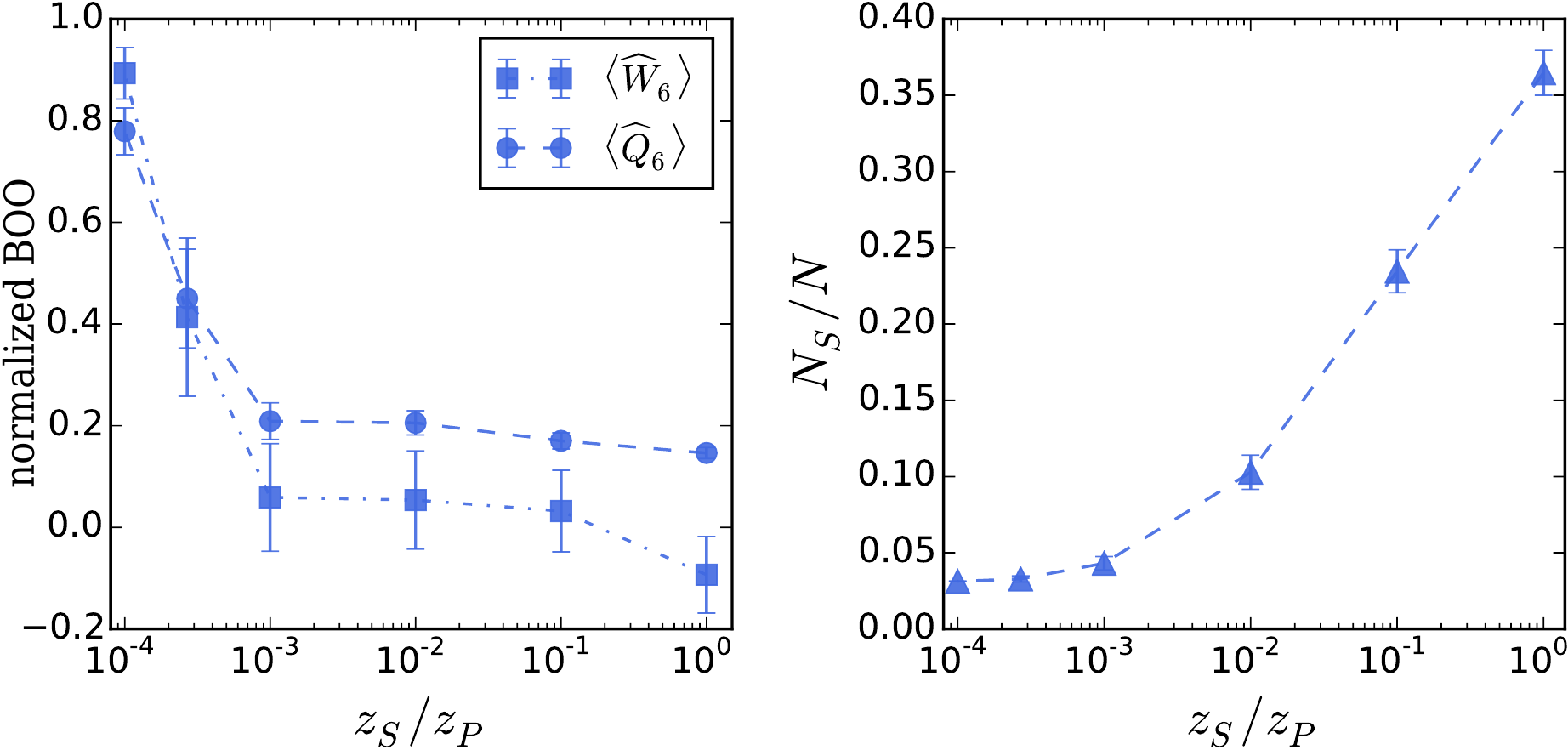}
  \caption{Averaged normalized BOOs (left) and the fraction of squares (right) are plotted as a function of $z_S/z_P$ for a perturbed system at $kT=0.15 \epsilon$ and $R=1.603\sigma$. \mm{Error bars represent two standard deviations away from the mean value obtained from 40 independent simulations.}}
  \label{fig:avgs_zs}
\end{figure}

For the ideal system, the projected landscapes and the average number of pentagons as a function of $\widehat{W}_6$ and $\widehat{Q}_6$ are illustrated in Fig.\,\ref{fig:perturb} with solid lines. At $kT=0.15\epsilon$, the lowest-energy packings are icosahedrally ordered with exactly 12 pentagons. Packings \textbf{i}, \textbf{ii}, \textbf{iii} are the three most probable packings ( \ie the lowest minima in the free-energy landscape) that the system samples at this temperature, with a high level of icosahedral order reflected by $\widehat{W}_6 \approx 1$. While \textbf{i} represents a perfect icosahedral arrangement, \textbf{ii} and \textbf{iii} are packings with 13 or 14 pentagons, respectively. These are packings in which one or two hexagons are replaced with pentagons, resulting in an increase in the average energy per particle, and also smaller normalized BOOs. At even lower temperatures, the icosahedral basin becomes the dominant global free-energy minimum, in agreement with previous studies \cite{Zandi_pnas,Paquay}. Similarly, the peaks in the free-energy landscape as a function of $\widehat{Q}_6$ include conformations in which a pentagon is replaced with a bigger hexagon; this substitution is very unfavorable energetically. In larger packings, the oscillations in the icosahedrally ordered region of the landscape shift to the right and become smaller in amplitude (see SI). This is because, in large packings, small movements of a larger number of particles can help to accommodate those substitutions more easily. It has been shown in many viral-capsid assembly studies \cite{Ceres_bc2002, Zlotnick2003, Hagan_review} that the subunits' binding energy is on the order $\approx 6\,kT$. Therefore, $kT=0.15 \epsilon$ is expected to be roughly room temperature when we interpret the results of our minimal model for virus capsids.

\begin{figure}
  \centering
  \includegraphics[width=1.0\linewidth]{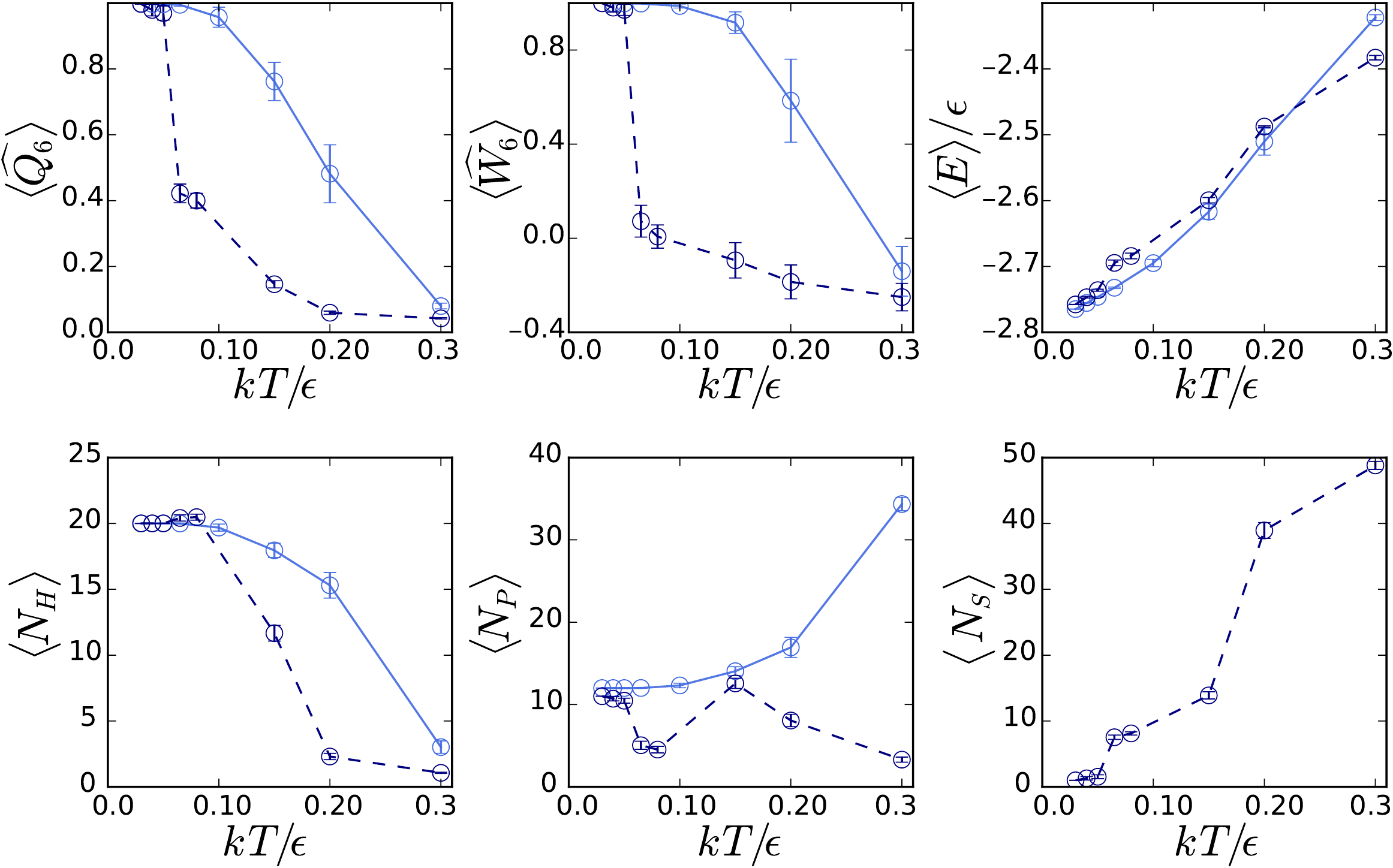}
  \caption{Order-disorder transition in the polygon-level CG model. Top panel plots display average normalized BOOs and the energy as a function of temperature for the ideal packing (solid lines) and for the perturbed packing (dashed lines) with $z_S/z_P = 1$ and $R=1.603\sigma$.  Average number of species as a function of temperature are also plotted in the bottom panel plots. \mm{Error bars represent two standard deviations away from the mean value obtained from 40 (20 for the ideal system) independent simulations.}}
  \label{fig:melting}
\end{figure} 

To investigate the stability of icosahedral packing in a perturbed system, we also considered a packing composed of hexagons, pentagons, and squares. Our umbrella-sampling results for the perturbed systems are shown with dashed, dotted and dashed-dotted lines, in the ascending order of the strength of the perturbation respectively, in Fig.\,\ref{fig:perturb}. For all cases (\ie for $z_S/z_P \geq 10^{-2}$), we observed that the existence of squares significantly changes the free-energy landscape of the system by shifting the most stable packing to the disordered region with small $\widehat{Q}_6$ and $|\widehat{W}_6|$ (packing \textbf{v} in Fig.\,\ref{fig:perturb}). It is interesting to note that even at $z_S/z_P = 10^{-3}$, the icosahedral order is effectively removed while only less than $5\%$ of particles are squares (see Fig.\,\ref{fig:avgs_zs}). This shows that in the presence of structural defects (squares), the icosahedral arrangement becomes unstable. Moreover, other energetically favorable symmetric configurations with a smaller symmetry group (\eg the packing in Fig.\,S6
-\textbf{D} with a $D_{5h}$ symmetry that is also responsible for the drop of $E$ when $\widehat{Q}_6 \to 0$), could be adopted more easily in the perturbed system \cite{Paquay}. Comparison between the relative population of squares in the CC-level CG simulations of the SAGEs and the number of squares in our perturbed packings suggests that for the SAGEs $z_S/z_P \approx 10^{-1}$. \mm{We note that the flexibility of the protein building block in the SAGE determines the relative activity of squares and we expect it to vary for different protein cages.} We also observed, in the disordered region of the landscape with increasing temperature, an increase in the number of non-hexagonal particles, which is also accompanied by a slower increase in the total number of particles. A qualitatively similar behavior was also observed for larger packings (see SI). 

To quantify the order-disorder transition in our packings we measured the average BOO parameters and the number of species as a function of temperature. The results are illustrated in Fig.\,\ref{fig:melting} for the ideal and perturbed packings. We observed a rapid increase in the total number of defects (extra pentagons and squares) at high temperatures. Upon introducing defects (squares) into the packing, the midpoint of the transition (\ie the melting point) shifted to lower temperatures for both $\widehat{Q}_6$ and $\widehat{W}_6$ transitions, indicating that the icosahedrally ordered packings became less stable for those perturbed systems. Moreover, the melting curves were characterized by a sharper transition in those perturbed packings. This is probably due to the fact that, starting from a disordered configuration in the perturbed packings, significantly larger number of particles are required to rearrange or change type cooperatively to reach the icosahedral arrangement. Therefore, for the perturbed packings, the order-disorder transition is characterized by a larger level of cooperativity compared to the ideal packing.

\section*{Discussion} Many natural systems including viruses have evolved over millions of years to form icosahedral shells from protein building blocks \cite{caspar1962physical,Zandi_pnas}. These provide the motivation for the design of synthetic self-assembling cages \cite{Fletcher:2013fx, King:2014cf, Indelicato:2016bf}. Natural icosahedral protein cages are formed usually via a hierarchical self-assembly process of highly specific building blocks with at least two stages: first two classes of mesoscale units combine, which then self-assemble further to a highly mono-dispersed regular shells. Using models on multiple scales, we have investigated the self-assembly of quasi-spherical shells from simple elementary building blocks, focusing on the regularity of their packing on the surface of the spherical cage. Using building blocks, that are able to pack with icosahedral symmetry, we examine how stable these highly symmetric structures are to perturbations that may arise from the flexibility of the interacting blocks. 
This allows us to explore the rather simpler self-assembly processes involving more flexible units that allow the formation of a variety of mesoscale objects and permitting structural defects. We find that they can also form shell-like structures with few or no holes, however with more variability in both the local structure on the shell and the overall symmetry and size of the shells formed. In particular, we find that by introducing a small number of structural defects, icosahedral packings are not the most stable structures for a wide range of parameters and that rather a disordered structure is found on the shell's surface. For many applications in protein design, however, such a variability is not a handicap and this suggests that icosahedral packings need not necessarily be the aim for synthetic design of micro-scale containers for packaging and transport. Indeed there are a number of protein shells (\eg the HIV virus capsid \cite{HIV1,HIVgag_defects}) that show disordered non-icosahedral surfaces, which seem also to be correlated with their greater variability. 
Our results are also relevant to colloidosomes that are fabricated by the self-assembly of colloidal particles onto the interface of emulsion droplets \cite{Dinsmore_Colloidosomes, Meng_Colloidosomes}. 
Our work provides a new design paradigm: we propose that by modulating the flexibilities of the components, one can control the regularity of the packing and, consequently, the surface properties of a synthetic cage.

\section*{Materials and Methods}

\subsection*{CC-level CG model of SAGEs}
We designed a CG model on the level of CCs to directly simulation the assembly of SAGEs. Each CC in this model is a rigid body with Lennard-Jones (LJ) attractive patches on its surface that drive formation of CC dimers and trimers \cite{Glotzer_nanoletter}. A permanent bond also connects a trimeric CC to a dimeric CC. Details of the model and its parameterization are explained in the SI.

\subsection*{Polygon-level CG model}
In this model, a polygon with $n$ edges is described as an LJ particles of diameter $\sigma_{\alpha} = \sigma /(2\sin(\pi/$n$))$. The pair-wise LJ interaction between polygons $\alpha$ and $\beta$, separated by a distance $r$ is described by 
$$ V_{\rm LJ}(r)=4\epsilon \big[  (\sigma_{\alpha\beta}/r)^{12} - (\sigma_{\alpha\beta}/r)^6  \big] $$
where, $\sigma_{\alpha \beta} = (\sigma_{\alpha} + \sigma_{\beta})/2$. The interaction is truncated at a cut-off distance $1.5 \sigma_{\alpha\beta}$. \mm{ We performed grand canonical MC simulations \cite{GC_MC} at the temperature $T$ and activities $\{z_\alpha\} = \{\exp(\mu_\alpha/kT)\}$, where $\mu_\alpha$ is the chemical potential of species $\alpha$, and $k$ is the Boltzmann constant. All particles are restricted to move on the surface of a sphere of radius $R$, and are allowed to change type. MC trial moves include (1) single particle moves, (2) deletion moves, (3) addition moves, (4) type-change moves and (5) position-swap of two dissimilar particles.} Care has been exercised to assure that MC moves satisfy the detailed balance condition. We performed extensive MC simulations for two system sizes $R=1.603\sigma$ and $R=2.469\sigma$, corresponding to icosahedral packings with $N=32$ and $N=72$ particles, respectively. We confirmed that these arrangements are energetically optimized configurations for the ideal system as $T\to0$.

\subsection*{Bond orientational order parameters}
We adopt the BOO of Steinhardt {\textit{et al.}} \cite{steinhardt_BOO} of $l$-fold symmetry to characterize our packing by considering the complex vector 
$$ Q_{lm}  = \frac{1}{N - N_H} \sum_{i=1}^{N-N_H} {Y_{lm}(\mathbf{r}_i)}~,$$
where the $Y_{lm}$ are spherical harmonics and $\mathbf{r}_i$ is the position of the non-hexagonal particle $i$ relative to the center of the sphere. One can construct the following rotational invariants
\begin{equation}
 Q_l = \left (\frac{4\pi}{2l+1} \sum_{m=-l}^{l} |Q_{lm}|^2 \right )^{1/2},
\end{equation}
\begin{equation} \label{Wl}
 W_l = \frac{
 \sum\limits_{\substack{m1,m2,m3 \\ m_1+m_2+m_3=0 }}{}
 \left[ \begin{smallmatrix}
    l & l & l  \\
    &&\\
    m_1 & m_2 & m_3 
  \end{smallmatrix}\right]
  Q_{lm_1}Q_{lm_2}Q_{lm_3}
  }
 {\left ( \sum_{m=-l}^{l} |Q_{lm}|^2 \right )^{3/2}}~,
\end{equation}
where the bracket in the third order invariant in Eq.\,\ref{Wl} represents the Wigner 3-$j$ symbol. It has been shown that these invariants can effectively be used to distinguish various types of local order within glasses and liquids \cite{steinhardt_BOO, lechner}. For our purpose, which is to distinguish only the icosahedral order, it is sufficient to restrict our consideration to $l=6$. In particular, $W_6$ is very sensitive to the level of icosahedral order within the packing \cite{steinhardt_BOO}; the more negative the $W_6$, the higher the level of icosahedral order.

\acknowledgements{The computational resources of the University of Bristol Advanced Computing Research Centre, and the BrisSynBio HPC facility are gratefully acknowledged. MM, DKS, RBS, DNW and TBL are supported by BrisSynBio, a BBSRC/EPSRC Synthetic Biology Research Center (BB/L01386X/1). RBS and DNW are funded by a BBSRC LoLa grant (BB/M002969/1). DNW is a Royal Society Wolfson Research Merit Award holder (WM140008).}



%

\end{document}